\documentclass[10pt]{article}

\usepackage{amsmath}
\usepackage{amssymb}

\usepackage{graphicx}

\usepackage{cite}

\usepackage[labelfont=bf,labelsep=period,justification=raggedright]{caption}

\makeatletter
\renewcommand{\@biblabel}[1]{\quad#1.}
\makeatother

\date{}

\usepackage{mathbbol}
\usepackage{color} 
\usepackage{ulem}
\usepackage{url}

\usepackage{color}

\begin{document}
\begin{flushleft}
{\Large
\textbf{DTW-MIC Coexpression Networks from Time-Course Data}
}
\\
Samantha Riccadonna$^{1}$,\\
Giuseppe Jurman$^{2,\ast}$,\\
Roberto Visintainer$^{2}$,\\
Michele Filosi$^{2}$,\\
Cesare Furlanello$^{2}$
\\
\bf{1} Research and Innovation Centre, Fondazione Edmund Mach, San Michele all'Adige, Italy 
\\
\bf{2} Fondazione Bruno Kessler, Trento, Italy 
\\
$\ast$ E-mail: jurman@fbk.eu\\
\end{flushleft}
\section*{Abstract}
When modeling coexpression networks from high-throughput time course data, Pearson Correlation Coefficient (PCC) is one of the most effective and popular similarity functions.
However, its reliability is limited since it cannot capture non-linear interactions and time shifts.
Here we propose to overcome these two issues by employing a novel similarity function, Dynamic Time Warping Maximal Information Coefficient (DTW-MIC), combining a measure taking care of functional interactions of signals (MIC) and a measure identifying horizontal displacements (DTW).
By using the Hamming-Ipsen-Mikhailov (HIM) metric to quantify network differences, the effectiveness of the DTW-MIC approach is demonstrated on both synthetic and transcriptomic datasets. 
\section*{Introduction}
Inferring a biological graph (\textit{e.g.}, a Gene Regulatory Network) from high-throughput longitudinal measurements of its nodes is nowadays one of the major challenges in computational biology, and several are the proposed solutions to this still unanswered question \cite{desmet10advantages,marbach10revealing,szederkenyi11inference}.
Although the problem is strongly non-linear, a simple but widespread solution such as the coexpression networks via correlation measures provides a good approximation \cite{allen12comparing,lopez13biostatistical,rotival14leveraging}, even outperforming more complex approaches \cite{allen12comparing,song12comparison,madhamshettiwar12gene,baralla09inferring}.
This follows from the consideration that functionally related genes share similar expression patterns \cite{lee04coexpression}, yielding that coexpression and functional relationships are correlated \cite{lavi12network,rapaport07classification,jansen02relating}.
Pearson Correlation Coefficient (PCC) is the most used measure \cite{zhang05general,langfelder08wgcna,horvath11weighted}, although alternative correlation functions can be also employed \cite{kumari12evaluation,kurt14comprehensive,dempsey11novel}.
However, PCC lacks sensitivity in case of non-linear relations and time shift between signals, thus the reliability of a coexepression network would benefit from a measure taking care of these characteristics.

Two measures proved to be effective in tackling these two issues: the Maximal Information Coefficient (MIC) and the Dynamic Time Warping (DTW).
MIC is a recent association measure based on mutual informations, aimed at detecting functional (linear and non-linear) dependencies between two variables \cite{reshef11detecting,speed11correlation,albanese12minerva,nature12finding}.
DTW is a classical measure evaluating the distance between two temporal sequences possibily varying in time or speed, applied to temporal sequences of video, audio, graphics and omics data \cite{sakoe78dynamic,keogh98enhanced,keogh00scaling,aach01aligning,furlanello06combining}.
Moreover, corrections have been proposed to DTW to overcome occuring drawbacks or to further refine the original measure \cite{keogh01derivative,rakthanmanon12searching,batista14cid}: however, the robustness of the original DTW is still acknowledged \cite{li14online}.
Finally, alternative measures to DTW do exist (see for instance \cite{elbakry10inference}), but they have not been extensively tested as DTW.

Here we propose to infer coexpression networks from -omics time series data by using as the similarity measure the function DTW-MIC, defined as the root mean square of MIC and of the similarity measure naturally induced by DTW, as an alternative to PCC.
To rate the reconstruction performance of the novel measure we evaluate the Hamming-Ipsen-Mikhailov (HIM) distance \cite{filosi14stability,jurman14him} of the inferred network from the gold standard in four cases where the gold standard is known.
In detail, we demonstrate DTW-MIC on four synthetic datasets generated by GeneNetWeaver (GNW) \cite{shaffter11genenetweaver} following the guidelines of the DREAM Challenge \cite{prill10towards}, and a transcriptomic dataset on the expression response of human T cells to phorbol 12-myristate 13-acetate (PMA) and ionomycin treatment \cite{rangel04modeling}, showing a consistent improvement in reconstruction over PCC.

The DTW-MIC measure, together with other association functions and the HIM distance, is included in the Open Source Web framework ReNette \cite{filosi14renette} for differential network analysis and visualization and in its R \cite{r14manual} package \textit{nettools}, available on the CRAN \url{http://cran.r-project.org} archive and on the GitHub repository \url{https://github.com/MPBA/nettools.git}.
\section*{Methods}
\subsection*{Time series similarity measures}
\subsubsection*{Maximal Information Coefficient}
The Maximal Information Coefficient (MIC) measure is a component of the Maximal Information-based Nonparametric Exploration (MINE) family of statistics, introduced in \cite{reshef11detecting,speed11correlation,nature12finding} for the exploration of two-variable relationships in multidimensional data sets.
Operatively, the MIC value is obtained by builiding several grids at different resolutions on the scatterplot of the two variables, then computing the largest possible mutual information achievable by any grid applied to the data, and finally normalizing to the $[0,1]$ range, where larger values correspond to higher similarity.
The two distinctive features of MIC are generality, \textit{i.e.}, the ability of capturing variable relationships of different nature, and equitability, that is the property of penalizing similar levels of noise in the same way, regardless of the nature of the relation between the variables. 
MIC (and the other MINE statistics) can be computed in R \cite{r14manual} by using the \textit{minerva} package \cite{albanese12minerva}.

To demonstrate the difference between PCC and MIC in detecting non-linear relationships between two variables, we introduce a simple synthetic example $\mathcal{E}_1$.
Consider the following five time series with 100 time points $\{t_i=i\colon 1\leq i\leq 100\}$:
\begin{displaymath}
\begin{split}
\label{eq:pcc_mic}
A(i) &= 0.01 i\\
B(i) &= \log_{100} i\\
C(i) &= 0.01i+\varepsilon(0.002i),\quad \varepsilon(z)\in\mathcal{U}(-z,z)\\
D(i) &= 0.5 \cos\log i  +0.65\\
E(i) &= 
\begin{cases} 
0 & \textrm{for $50\leq i\leq 70$} \\
D(i)-0.15 & \textrm{otherwise}\ ,
\end{cases}
\end{split}
\end{displaymath}
where $\mathcal{U}(a,b)$ is the uniform distribution with extremes $a<b$.

\begin{figure}[!ht]
\begin{center}
\includegraphics[width=\textwidth]{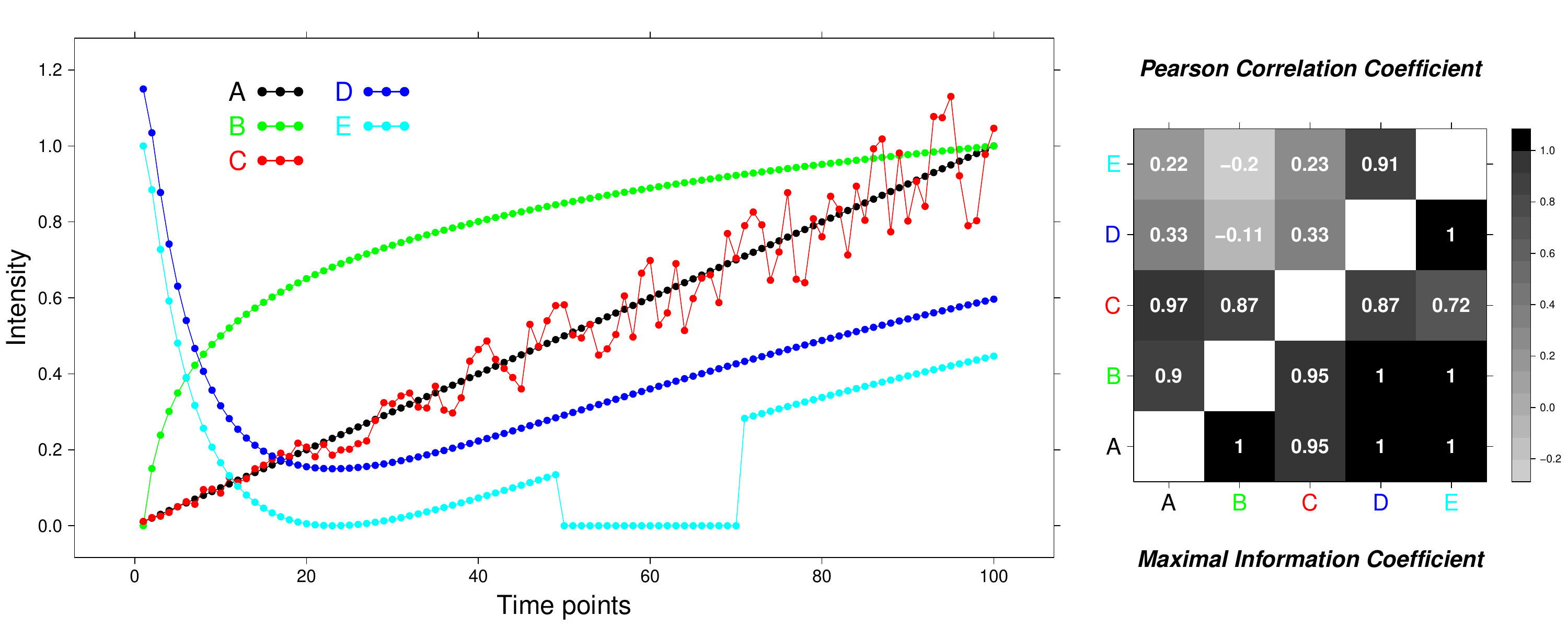}
\end{center}
\caption{\textbf{Example $\mathcal{E}_1$:} PCC versus MIC in a synthetic example with five time series $A$--$E$ on 100 time points (left) and the corresponding PCC values (right panel, top-left triangle) and MIC values (right panel, bottom-left triangle) for all pairs of time series.}
\label{fig:pcc_mic}
\end{figure}

While $A(i)$ is just $1/100-$th of the identity map, $B(i)$ is a logarithmic map, $C(i)$ is obtained by $A(i)$ by adding a 20\% level of uniform noise, $D(i)$ is a more complex non-linear map merging a trigonometric and a logarithmic relation and, finally, $E(i)$ is obtained by $D(i)$ by a vertical offset and then flattening to zero all the values in the time interval [50,70].
In Fig.~\ref{fig:pcc_mic} the plot of the five time series $A$-$E$ is displayed together with the values of the Pearson Correlation Coefficient and the Maximal Information Coefficient for all pairs of sequences.
As predictable, MIC is able to capture the functional relationship linking all pairs of time series, even in presence of a moderate level of noise: all MIC values are larger than 0.72, and in six cases out of ten MIC attains the upper bound 1.
On the other hand, PCC is close to one only when evaluating the pairs $(A,B)$, $(A,C)$, $(B,C)$ and $(D,E)$, while all the remaining six cases display a correlation score smaller than 0.33, supporting the uneffectiveness of PCC as a similarity measure for complex longitudinal data.
As a relevant example, note that $B(i)$ has a strong functional dependence with $E(i)$ and $F(i)$ although the shape of the corresponding curves are hugely different: this behaviour is well captured by the two measures, attributing MIC similarity 1 to both $(B,D)$ and $(B,E)$, while the corresponding values for PCC are negative.

\subsubsection*{Dynamic Time Warping}
The Dynamic Time Warping (DTW) \cite{keogh98enhanced,keogh00scaling} is a measure of distance between two sequences taking care of occurring time shifts between the series.
In particular, the DTW algorithm finds an optimal match between the two given series by non-linearly warping them in the time dimension to determine a measure of their dissimilarity, stretching (or compressing) the time axis: for a comprehensive reference, the reader is referred to \cite{gusfield97algorithms}.
Thus, by definition, the shapes of the compared curves become a more crucial factor in DTW rather than the pointwise distance of the time series values.
To obtain a similarity measure $\textrm{DTW}_s$ from the distance DTW we use the function $\textrm{DTW}_s=1/(1+\textrm{DTW}_d)$, where $\textrm{DTW}_d$ is the normalized distance between two series, as computed in the R \cite{r14manual} package \textit{dtw} \cite{giorgino09computing}.

In what follows, a synthetic example $\mathcal{E}_2$ is used to highlight the difference between DTW and PCC for increasing time shift, with and without a moderate noise level. 
This example mimicks a common situation in -omics data, when the activation of a gene induces a delayed activation of a previously inactive gene $h$, with a similar expression level curve, affected by a certain amount of noise.

\begin{figure}[!ht]
\begin{center}
\includegraphics[height=0.95\textheight]{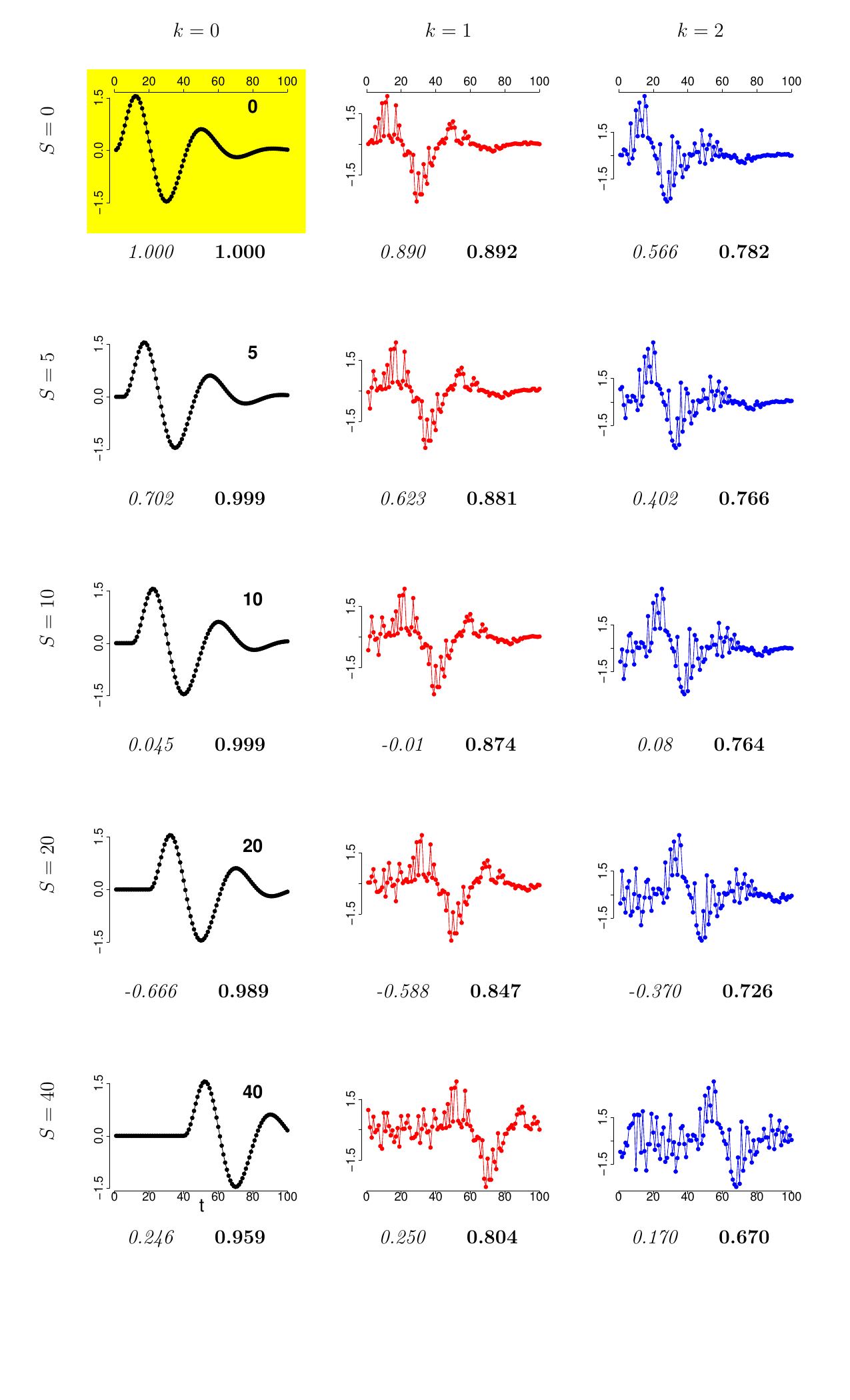}
\end{center}
\caption{\textbf{Example $\mathcal{E}_2$:} PCC and $\textrm{DTW}_s$ versus the reference series $r$ for the 15 time series $\{r_s^{[k]}\}$ with $s=0,5,10,20,40$ and $k=0,1,2$. Each row corresponds to a different value of $s$, indicated by the figure in the top right corner of the plot in the first column. Each column corresponds to a different value of $k$: 0 on the left, with black curves, 1 in the centre, with blue curves and 2 on the right, with red curves. The plot in the top left panel with yellow background is the reference time series $r_0^{[0]}=r$. Under each panel, the corresponding values are reported for $P(s,k)$ (italic) and $D(s,k)$ (boldface).
}
\label{fig:pcc_dtw}
\end{figure}

\begin{figure}[!ht]
\begin{center}
\includegraphics[width=\textwidth]{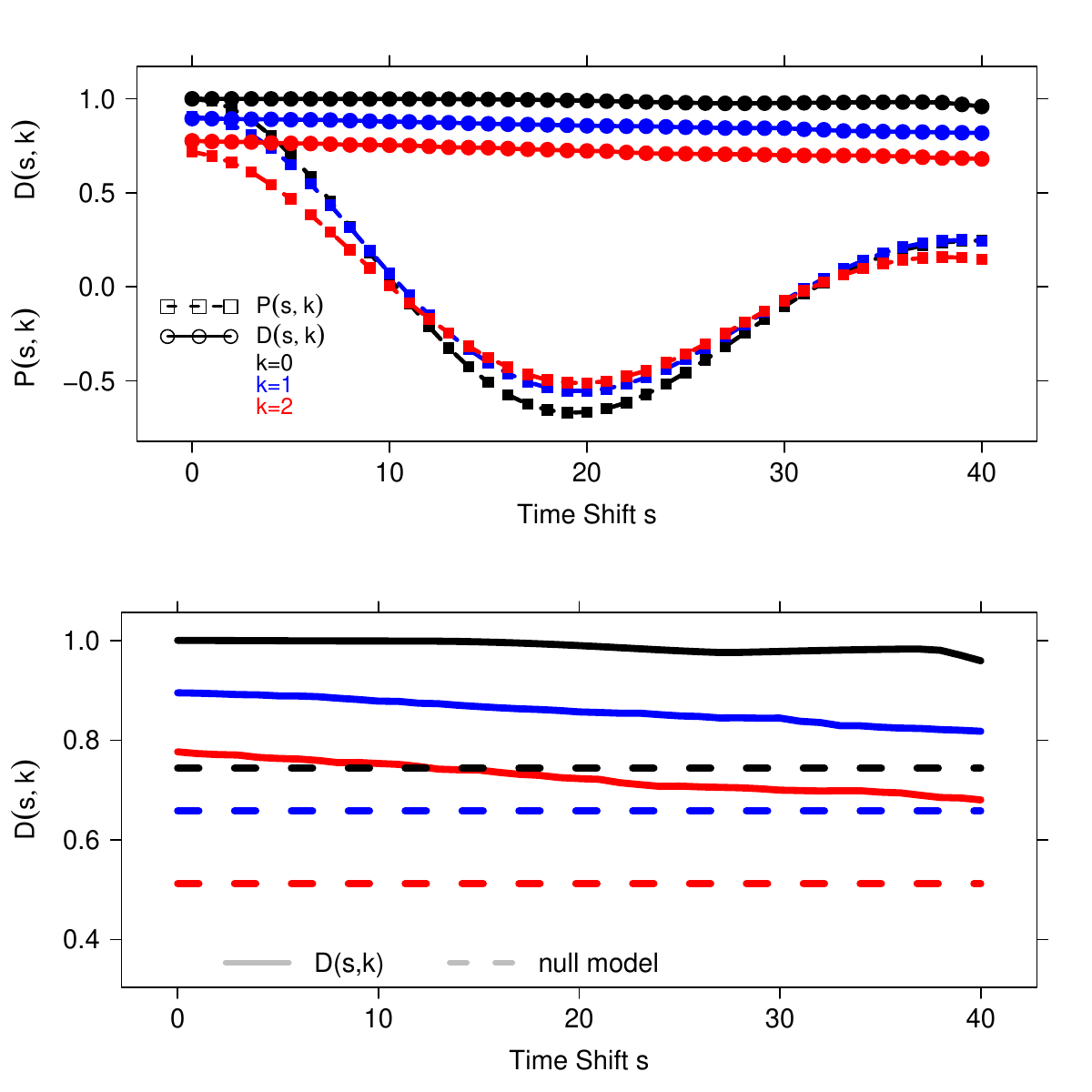}
\end{center}
\caption{\textbf{Example $\mathcal{E}_2$:} PCC and $\textrm{DTW}_s$ versus the reference series $r$ for the $\{r_s^{[k]}\}$ with $k=0,1,2$ when the time shift $s$ ranges between 0 and 40. Squares correspond to $P(s,k)$, while circles and solid lines indicate $D(s,k)$; the different noise levels $k=0,1,2$ are denoted by curves in black, blue and red respectively. The dashed lines in the bottom panel indicate the no information value for $\textrm{DTW}_s$ as obtained by the null model described in the main text.}
\label{fig:pcc_dtw_all}
\end{figure}

Consider the time series $\{r(i)\colon 1\leq i \leq 100\}$ on 100 time points defined as follows:
\begin{displaymath}
\begin{split}
r\colon & [1,100]\cap\mathbb{N} \to \mathbf{R} \\
& i \mapsto  \frac{1}{10}  \mathrm{e}^{-\frac{2}{25}i} \sqrt{i^3} \sin\left(\frac{3}{20}i\right)\ ,
\end{split}
\end{displaymath}
whose graph is displayed in the top-left panel (with yellow background) of Fig.~\ref{fig:pcc_dtw}.
Moreover, define the following family of time series originated by $r$:
\begin{displaymath}
\begin{split}
r_s^{[k]}\colon& [1,100]\cap\mathbb{N} \to \mathbf{R} \\
& i \mapsto 
\begin{cases}
\varepsilon(k) & \textrm{for $i\leq s$}\\
r(i-s)+\varepsilon(k\cdot r(i-s))  & \textrm{for $s<i\leq 100$}\ ,
\end{cases}
\textrm{ where } \varepsilon(z)\in\mathcal{U}(-z,z)\ .
\end{split}
\end{displaymath}
In this notation, $r_0^{[0]}(i)=r(i)$.
Finally, define the two functions
\begin{align*}
P\colon & \mathbb{N}\times\mathbb{R}_0^+ \to [-1,1] & D\colon & \mathbb{N}\times\mathbb{R}_0^+ \to [0,1] \\
& (s,k) \mapsto \textrm{PCC}(r_s^{[k]},r) 
&
& (s,k) \mapsto \textrm{DTW}_s(r_s^{[k]},r) 
\end{align*}
In Fig.~\ref{fig:pcc_dtw} the plots of the 15 time series $\{r_s^k\colon s=0,5,10,20,40\ ,k=0,1,2\}$ is shown, together with the corresponding values of $P(s,k)$ (italic) and $D(s,k)$ (boldface).
Moreover, in the top panel of Fig.~\ref{fig:pcc_dtw_all} the curves $P(s,k)$ (squares) and $D(s,k)$ (dots) are displayed for $k=0,1,2$ (in black, blue and red respectively) versus the time shift $s$ ranging from 0 to 40.
From both figures it is possible to infer the effectiveness of DTW in capturing the dependence between $r(s,k)$ and $r$, even for large time shift $s$ and high noise level $k$.
In particular, as a function of the time shift $s$, the value for DTW monotonically decreases from 1 to 0.959, 0.804, 0.670 for $k=0,1,2$ respectively, and $D(s,0)>D(s,1)>D(s,2)$ consistently along the whole range $0\leq s\leq 40$.
On the other hand, PCC has a much more erratic behaviour, rapidly decreasing to very low correlation level even for small time shifts $s>5$, and remaining under $PCC<0.3$ for all values $s>7$.
Furthermore, the curves for $P(s,k)$ corresponding to different noise level $k$ mutually intersecate, showing a non consistent behaviour of Pearson correlation in these case with respect to increasing noise.
Finally, to assess the significativity of the $D(s,k)$ values, a null model is built as follows: 
\begin{enumerate}
\item given two real values $m<M$, generate a set $\{ \eta_j\colon \eta_j\in [m,M]^{100}, 1\leq j \leq 2N, \eta_j(i)\in \mathcal{U}(m,M)\}$ of $2N$ random vectors $\eta_j$ on 100 time points with values randomly and uniformly sampled between $m$ and $M$; 
\item compute the set of distances $D_m^M = \{\textrm{DTW}_s(\eta_j, \eta_{j+N})\}$;
\item consider the distribution of $D_m^M$ as the null distribution.
\end{enumerate}
Here we use $N=1000$ and, given a noise level $k$, we define $m=\displaystyle{\min_{\substack{0\leq s\leq 40\\ 1\leq i\leq 100}} r_s^{[k]}(i)}$ and $M=\displaystyle{\max_{\substack{0\leq s\leq 40\\ 1\leq i\leq 100}} r_s^{[k]}(i)}$.
For all the three cases $k=0,1,2$, the distribution of the set $D_m^M$ is Gaussian shaped, and the 95\% Student bootstrap confidence intervals around the mean are quite narrow, namely (0.7429,0.7441), (0.6570,0.6584) and (0.5115,0.5130) for $k=0,1,2$ respectively. 
Thus the mean values $\overline{D_m^M}$, \textit{i.e.},  0.7435 ($k=0$), 0.6577 ($k=1$) and 0.5121 ($k=2$), can be used as significativity thresholds, as shown in the bottom panel of Fig.~\ref{fig:pcc_dtw_all}: in all the three cases, throughout the whole range $0\leq s\leq 40$, the curve $P(s,k)$ lies above the corresponding significativity threshold value.

\subsubsection*{DTW-MIC}
We introduce here DTW-MIC, a novel measure for evaluating the similarity between two time series defined as the root mean square of MIC and $\textrm{DTW}_s$:
\begin{displaymath}
\textrm{DTW-MIC}(T_1,T_2)=\frac{1}{\sqrt{2}}\sqrt{\text{DTW}_s(T_1,T_2)^2+\text{MIC}(T_1,T_2)^2}\ .
\end{displaymath}
By definition, DTW-MIC joins the contributions of both MIC and $\textrm{DTW}_s$, thus taking care of time shifts and non-linear functional relations.
This characteristic makes DTW-MIC more effective than PCC and than MIC and DTW considered separately, as demonstrated in the synthetic example $\mathcal{E}_3$ described hereafter.

\begin{figure}[!ht]
\begin{center}
\includegraphics[width=0.8\textwidth]{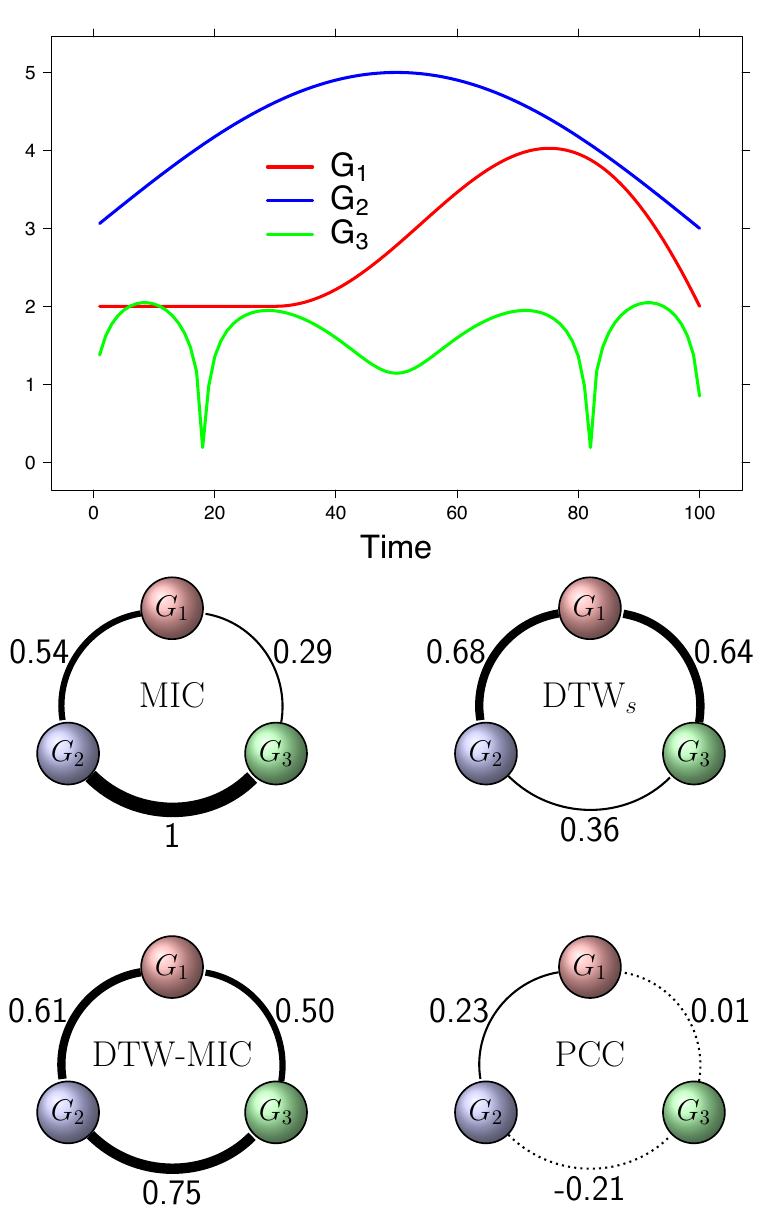}
\end{center}
\caption{\textbf{Example $\mathcal{E}_3$:} Plots (top) and PCC, MIC, $\textrm{DTW}_s$ and DTW-MIC weighted coexpression networks (bottom) for the set $\mathcal{G}$ of the three time series $G_1, G_2$ and $G_3$ (in red, blue and green respectively). Arc width is proportional to edge weight.}
\label{fig:dtwmic}
\end{figure}

Consider a set $\mathbf{g}$ of three genes $g_1$, $g_2$ and $g_3$ and the corresponding time series of expression $\mathcal{G}=\{G_1, G_2, G_3\}$ on 100 time points $1\leq i\leq 100$ defined as follows:
\begin{displaymath}
\begin{split}
G_1(i) &= 
\begin{cases}
2 & \textrm{for $1\leq i\leq 30$} \\
2+\frac{i-30}{20}\sin\frac{(i-30)\pi}{70} & \textrm{for $31\leq i\leq 100$} \\
\end{cases}
\\
G_2(i) &= 3+2\sin\frac{i\pi}{100}\\
G_3(i) &= 2+\log\sqrt{\left |\frac{1}{10}+\sin 3 (G_2(i)-3)\right|}\ .
\end{split}
\end{displaymath}
The graph of the functions in $\mathcal{G}$ are plotted in the top panel of Fig.~\ref{fig:dtwmic}, while in the bottom panel we list all values $\{M(i,j)\colon M\in\{\textrm{PCC},\textrm{MIC},\textrm{DTW}_s,\textrm{DTW-MIC}\}\textrm{ and } 1\leq i< j\leq 3\}$ by representing, for each similarity measure $M$, the corresponding coexpression network on the set of nodes $\mathbf{g}$.
All the three pairs of series have a very low correlation ($\textrm{PCC}\leq 0.23$), while DTW-MIC is able to capture the existing relation between them ($\textrm{DTW-MIC}\leq 0.5$), even when these relations are of different nature.
In fact, $G_2$ and $G_3$ have a low DTW similarity, but a high MIC correlation, while the opposite happens for $G_1$ and $G_3$; finally, the last pair $G_1$, $G_2$ shares both a moderate MIC and DTW similarity value.
In all three cases the resulting DTW-MIC value is above the significativity threshold computed from the null model described in the previous section, which is 0.52 for $(G_1,G_2)$, 0.29 for $(G_2,G_3)$ and 0.39 for $(G_1,G_3)$.

\subsection*{Network Analysis}
\subsubsection*{Co-expression networks}
An effective method for simultaneously analysing the mutual relations among a group of interacting agents is provided by the graph theory and it consists in building the complex network having the agents as nodes and inferring the (weight of the) edges connecting them using some similarity measure between their signals.
A typical example in -omics science is represented by the gene networks, where the nodes are genes and an edge between two genes is weighted by the similarity between their expression levels in a time window as read by, for instance, microarray or sequencing technology (or, in case of a binary network, the edge is declared to exist only if the similarity value lies above a chosen threshold).
These graphs are called coexpression networks and their most widespread model is the Weighted Gene Co-expression Network Analysis (WGCNA) \cite{zhang05general,langfelder08wgcna,horvath11weighted} where the adopted similarity is the (absolute) PCC, soft thresholded by a power function. 
In detail, given $N$ genes and their expressions $g_1,\ldots, g_n$, the resulting WGCNA network is described by the adjacency matrix $A$ whose entries are defined as 
\begin{equation}\label{eq:wgcna}
a_{ij}=M(g_i,g_j)^\beta\ ,
\end{equation}
for $M=|\textrm{PCC}|$ and $\beta$ a positive power, usually tuned according to additional constrains, such as the scale-freeness \cite{desollaprice65networks,barabasi99emergence} of the network; the default choice in the WGCNA R/Bioconductor package \cite{langfelder08wgcna} is $\beta=6$.
In the Results section we will use the WGNCA framework with the novel DTW-MIC as the $M$ measure in Eq.~\ref{eq:wgcna}, comparing the obtained networks with those inferred by the classical choice $M=|\textrm{PCC}|$.

\subsubsection*{Hamming-Ipsen-Mikhailov distance}
For the quantitative assessment of the difference between two networks sharing the same nodes a graph distance is required.
Among all metrics described in the literature, we choose the Hamming-Ipsen-Mikhailov (HIM) distance for its consistency and robustness \cite{jurman10introduction,jurman14him}.
The HIM distance for network comparison is defined as the product metric of the Hamming distance H \cite{tun06metabolic,dougherty10validation} and the Ipsen-Mikhailov distance IM \cite{ipsen02evolutionary}, normalized by the factor $\sqrt{2}$ to set its upper bound to 1:
\begin{displaymath}
\textrm{HIM}(N_1,N_2)=\frac{1}{\sqrt{2}}\sqrt{\text{H}(N_1,N_2)^2+\text{IM}(N_1,N_2)^2}\ ,
\end{displaymath}
for $N_1,N_2$ two undirected (possibly weighted) networks.
The drawback of edit distances (such as H) is their locality, as they focus only on the network parts that are different in terms of presence or absence of matching links \cite{jurman10introduction}.  
Spectral distances like IM are global, since they take into account the whole graph structure, but they cannot distinguish isomorphic or isospectral graphs, which can correspond to quite different conditions within the biological context. 
The HIM distance is a solution tackling both issues: details on HIM and its two components H and IM together with a few application examples are given in \cite{jurman14him,filosi14stability}.
In particular, HIM distance can be computed also for directed networks by using an alternative description of the graph topology.
Values of HIM distance range from 0 -- when comparing identical networks -- to 1, attained only when comparing the full and the empty network.

\begin{figure}[!ht]
\begin{center}
\includegraphics[width=\textwidth]{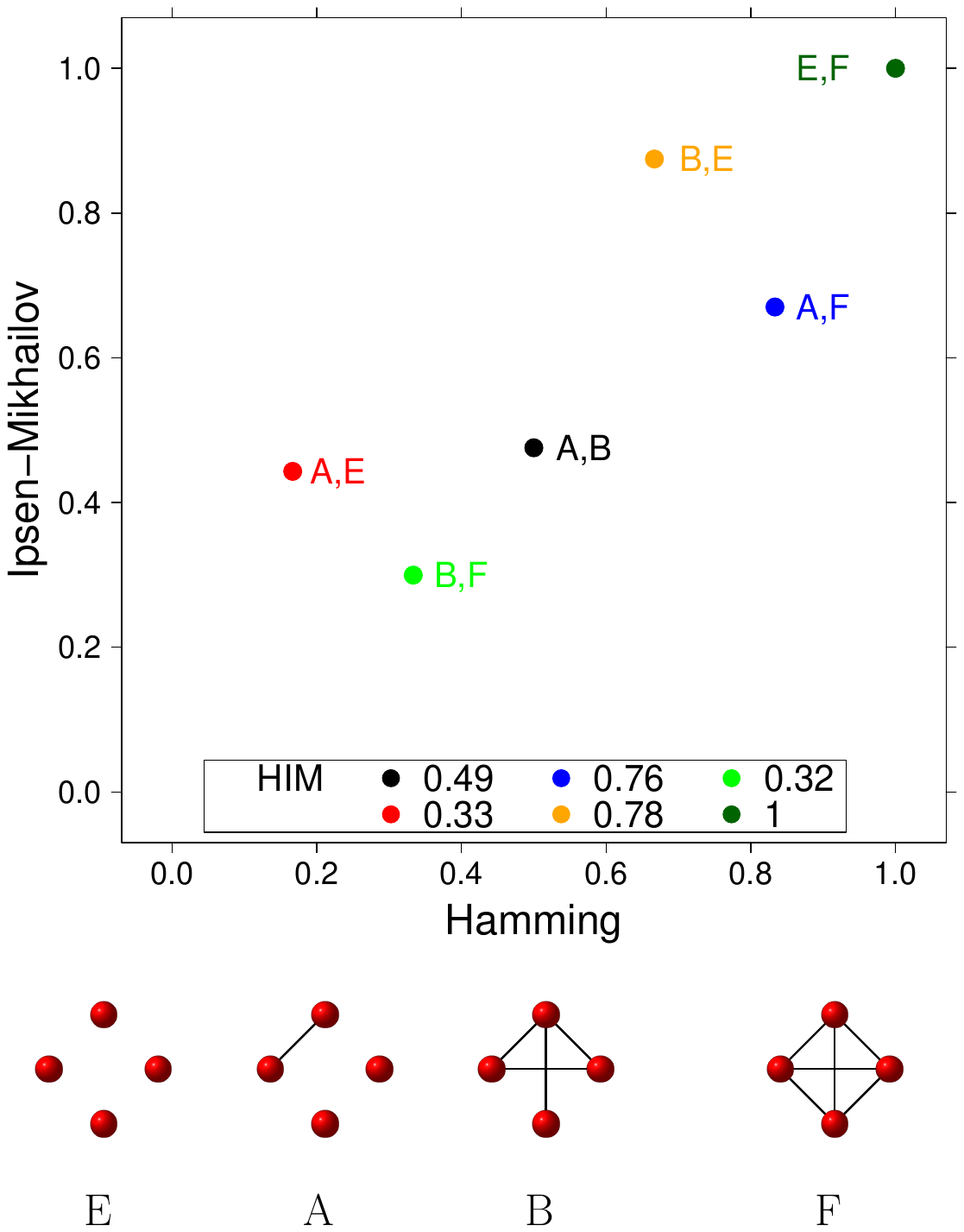}
\end{center}
\caption{\textbf{Example $\mathcal{E}_4$:} Mutual HIM distances in the Hamming $times$ Ipsen-Mikhailov space between 4 non-isospectral graphs A, B, E, F on 4 vertices, whose topology is shown below the plot. Distance values are listed in the plot legend.
}
\label{fig:him}
\end{figure}

In the example $\mathcal{E}_4$ shown in Fig.~\ref{fig:him}, we selected four non-isospectral networks on four vertices, namely the empty graph E, the full graph F, a network with 1 edge A and a network with 4 edges including a 3-cycle.
For these 4 graphs, the mutual H, IM and HIM distances are computed and reported as points on the H $\times$ IM plane, where each distance $\textrm{HIM}(P,Q)$ between two graphs $P$ and $Q$ is represented by a point of coordinates $R=(\textrm{H}(P,Q),\textrm{IM}(P,Q))$ and its HIM value is the length of the segment connecting $R$ to the origin $(0,0)$, divided by $\sqrt{2}$.
The visualization in the H $\times$ IM plane allows the relative comparison of the values of the two components of the distance: for instance, the Hamming distance between A and E is half the Hamming distance between B and F (1/6 vs. 1/3), but the IM component is much larger for the former pair, yielding two quite similar values for HIM. 

\section*{Results}
In this section we apply the novel DTW-MIC similarity measure to two case studies in life science.

In detail, in the first application a suite of three synthetic datasets is generated, inspired to real biological systems.
Each dataset includes a network $\mathcal{N}$ of connections between $n$ genes, together with the corresponding time series describing, for each gene, the dynamics of the expression level.
Aim of the study is the reconstruction of a network $\mathcal{N}'$ starting from the longitudinal data, and the comparison of the inferred network $\mathcal{N}$' with the ground truth $\mathcal{N}$ to evaluate the goodness of the inference algorithm.

The second task has the same goal, but expression level measurements come from a publicly available microarray dataset from a human cohort and the real network is unknown. 
Our strategy is the same in both applications and it includes two steps: first, the reconstruction of the network in the WGCNA framework in the classical approach via PCC and through the DTW-MIC, and then the evaluation of the HIM distance of the reconstructed networks from the true graph.

\subsection*{GeneNetWeaver Yeast \& \textit{E. coli} data} 
The datasets for the synthetic example are generated by GeneNetWeaver (GNW) \cite{marbach09generating,shaffter11genenetweaver} an open-source tool for in silico benchmark generation, available at the web address \url{http://gnw.sourceforge.net/genenetweaver.html}.
GNW generates realistic network structures of biologically plausible benchmarks by extracting modules from known gene networks of model organisms like yeast and \textrm{E. coli} \cite{shaffter12gnw}, endowing them with dynamics using a kinetic thermodynamical model of transcriptional regulation with added internal noise, allowing for different types of customizable perturbations.
According to the user prescribed constraints and given a chosen network topology, GNW can also produce (steady states and) time course datasets with the expression levels of the network nodes.
The annual Dialogue for Reverse Engineering Assessments and Methods (DREAM, \url{http://www.the-dream-project.org/}) Challenge \cite{stolovitzky07dialogue,stolovitzky09lessons,prill10towards,marbach10revealing,prill11crowdsourcing,marbach12wisdom} initiative for the quantitative comparison of network inference methods relies on GNW for the synthetic benchmark datasets.

Three synthetic networks are generated by GNW for the first application task, namely $\textrm{Yeast}_{20}$, $\textrm{Ecoli}_{20}$, $\textrm{Ecoli}_{50}$, where the name points to the original reference network and the subscript indicates the number of nodes.
In detail, $\textrm{Yeast}_{20}$ is a subnet of the Yeast transcriptional regulatory network with 4441 nodes and 12873 edges \cite{shaffter12gnw,balaji06comprehensive}, while $\textrm{Ecoli}_{20}$ and $\textrm{Ecoli}_{50}$ are subnets of the \textit{Escherichia coli} transcriptional regulatory network with 1502 nodes and 3587 edges, corresponding to the TF-gene interactions of RegulonDB release 6.7 of May 2010 \cite{shaffter12gnw,gamacastro08regulondb}.
In all cases, the selected genes are randomly extracted from the whole set of nodes only requiring that half of the selected nodes be regulators.

\begin{figure}[!ht]
\begin{center}
\includegraphics[width=\textwidth]{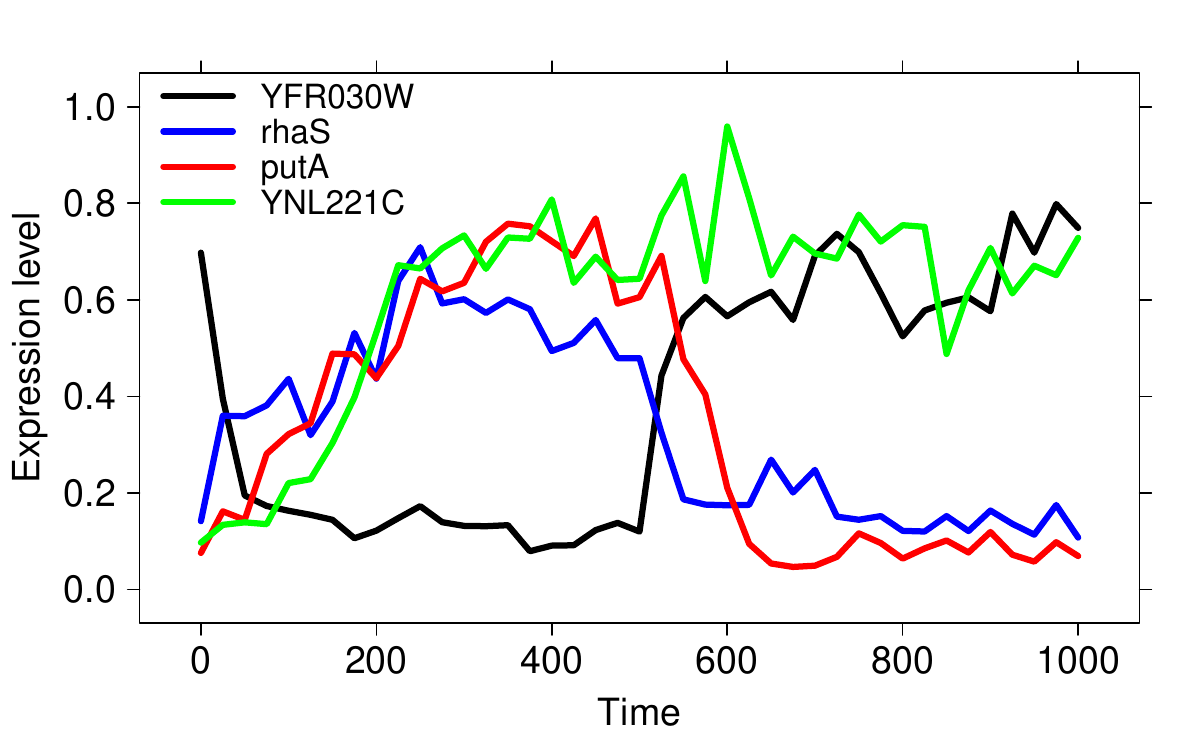}
\end{center}
\caption{\textbf{GeneNetWeaver time series:} examples of 4 longitudinal expression level data generated by the GNW kinetic model for the synthetic subgraph of Yeast and \textit{E. coli} regulatory networks. Time course data are defined on 41 time points $0,\ldots, 1000$ and they correspond to the genes YFR030W (black, from $\textrm{Yeast}_{20}$), YNL221C (green, from $\textrm{Yeast}_{20}$ with dual gene knockout), rhaS (from $\textrm{Ecoli}_{20}$) and putA (from $\textrm{Ecoli}_{50}$).
}
\label{fig:gnw_ts}
\end{figure}

For each network, 10 longitudinal datasets $\{d_1,\ldots,d_{10}\}$ of expression levels are generated by a dynamic model mixing ordinary and stochastic differential equations, on 41 time points equally spaced between time 0 and time 1000 $\{t_0=0,t_1=25,\ldots,t_{40}=1000\}$.
In each series, the initial time point $t_0=0$ corresponds to the wild-type steady-state and, from that moment onwards, a perturbation is applied until time point $t_{20}=500$: at that point, the perturbation is removed, and the gene expression level goes back from the perturbed to the wild-type state \cite{shaffter12gnw}.
Moreover, a moderate level of noise is added to all the datasets, namely 0.5\% for the Yeast data and 1\% for the E.coli data; in both cases, the selected model is the microarray noise model described in \cite{tu02quantitative}.
Both the noise model and the perturbation scheme are chosen according to the configuration of the DREAM4 challenge \cite{shaffter12gnw}.
As an example, in Fig.~\ref{fig:gnw_ts} we show the plots of the generated time course data of four genes belonging to the selected subnets $\textrm{Yeast}_{20}$, $\textrm{Ecoli}_{20}$ and $\textrm{Ecoli}_{50}$.

\begin{figure}[!ht]
\begin{center}
\includegraphics[width=0.8\textwidth]{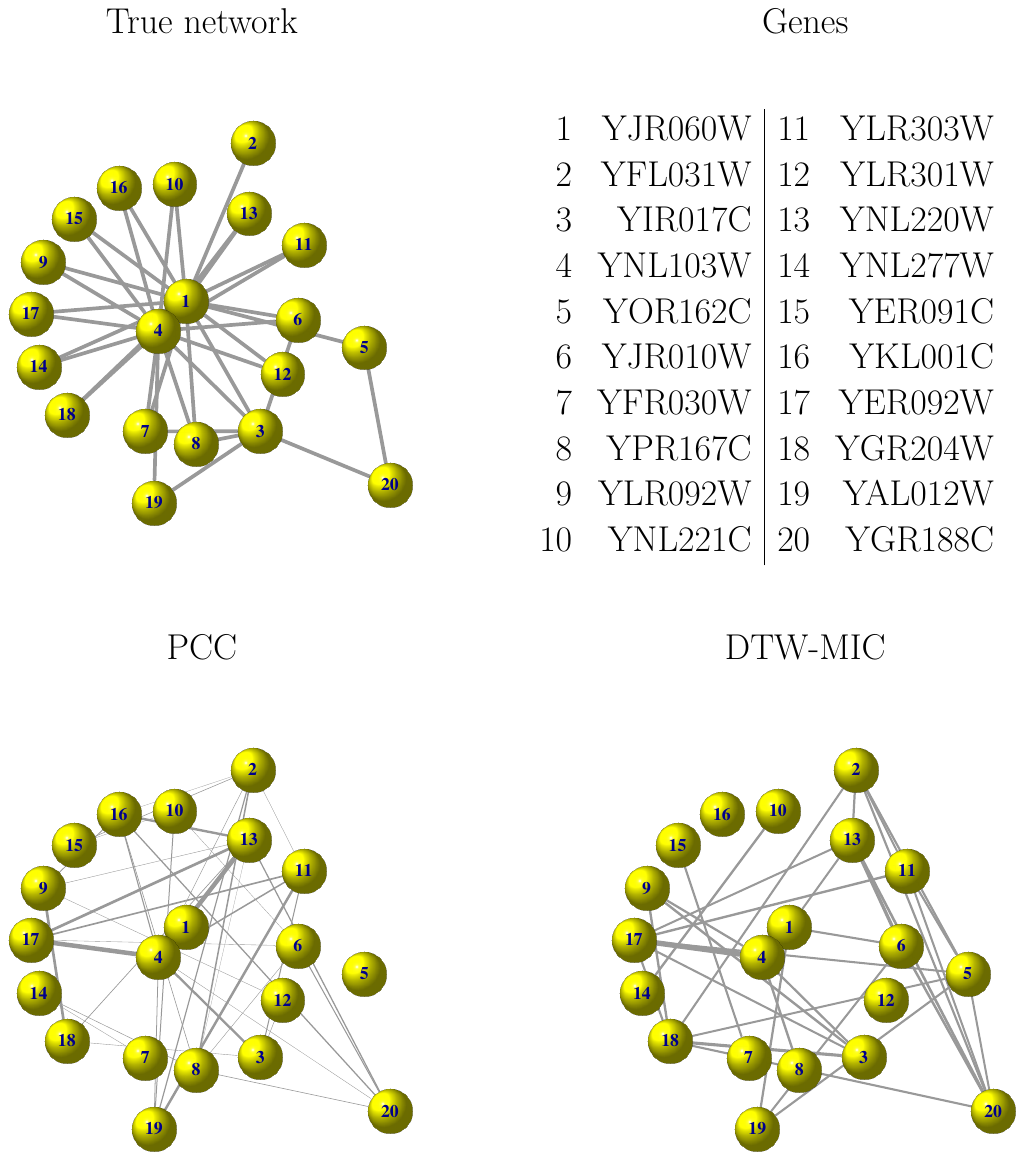}
\end{center}
\caption{\textbf{GeneNetWeaver data:} example of network reconstruction and comparison with ground truth. In the top panels, the topology of the synthetic true network $\textrm{Yeast}_{20}$ (top left) is shown together with the Systematic Name of its 20 genes (top right). In the two bottom panels, the network $\textrm{Yeast}_{20}$ as inferred from the time course dataset $d_1$ by PCC (bottom left) and DTW-MIC (bottom rright). For the reconstructed networks, edge width is proportional to arc weight; edges with smaller weights (threshold is 0.001 for PCC and 0.135 for DTW-MIC) are not drawn to avoid cluttering the image. Distance from the true network is 0.57 for the inference by PCC, and 0.23 for the reconstruction by DTW-MIC.
}
\label{fig:gnw_nets}
\end{figure}

\begin{figure}[!ht]
\begin{center}
\includegraphics[width=\textwidth]{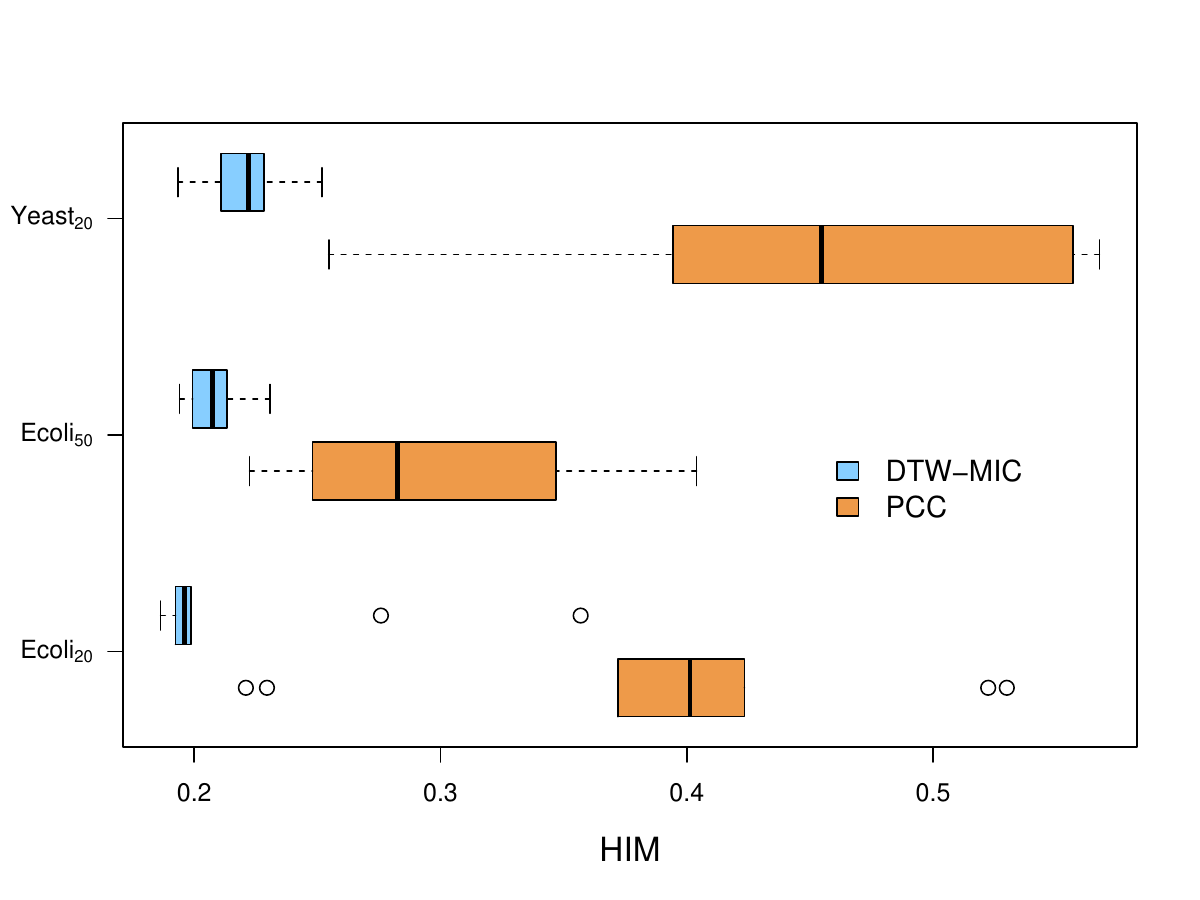}
\end{center}
\caption{\textbf{GeneNetWeaver data:} box and whisker plot of the HIM distance between the time series inferred networks and the true graphs, listed in Tab.~\ref{tab:gnw}. For each true network $\textrm{Yeast}_{20}$, $\textrm{Ecoli}_{20}$ and $\textrm{Ecoli}_{50}$, 10 different graphs are reconstructed by PCC and DTW-MIC similarity measures.
}
\label{fig:gnw_bw}
\end{figure}

In each of the three cases $\textrm{Yeast}_{20}$, $\textrm{Ecoli}_{20}$ and $\textrm{Ecoli}_{50}$, a network is inferred both by PCC and by DTW-MIC from each of the time course dataset $\{d_1,\ldots,d_{10}\}$, and the obtained graph is compared via the HIM distance to the corresponding ground truth network. 
As an example, in Fig.~\ref{fig:gnw_nets} we show the true $\textrm{Yeast}_{20}$ graph and the corresponding networks as reconstructed from the dataset $d_1$ by PCC and by DTW-MIC.
The results are reported in Table~\ref{tab:gnw} and summarized in the box and whisker plot of Fig.~\ref{fig:gnw_bw}.
The networks inferred by DTW-MIC are consistently closer to the true network than the PCC graphs, with also smaller standard deviation over the 10 experiments.

For the $\textrm{Yeast}_{20}$ dataset, 4 additional time course datasets were generated on the same timepoints, but with a dual gene knockout: the curve of gene YNL221C in Fig.∼\ref{fig:gnw_ts} is an example of the generated trajectories.
Again, the DTW-MIC inferred networks are closer to the gold standard than the PCC graphs: in the 4 experiments, the HIM distances for DTW-MIC are 0.23, 0.21, 0.24 and 0.21, while for PCC networks the corresponding values are 0.30,  0.27, 0.31 and 0.47 respectively.

\subsection*{Human T-cell data}
Rangel and colleagues in \cite{rangel04modeling} investigated the dynamics of the activation of T-lymphocites by analysing the response of the human Jurkat T-cell line subjected to a treatment with PMA and ioconomin.
Operatively, they measured the expression of 58 genes across 10 time points (0, 2, 4, 6, 8, 18, 24, 32, 48, and 72 hours after treatment) with two series of respectively 34 and 10 replicates on a custom microarray built by spotting PCR products on amino-modified glass slides using a Microgrid II spotter.

\begin{figure}[!ht]
\begin{center}
\raisebox{3cm}{\includegraphics[width=.7\textwidth]{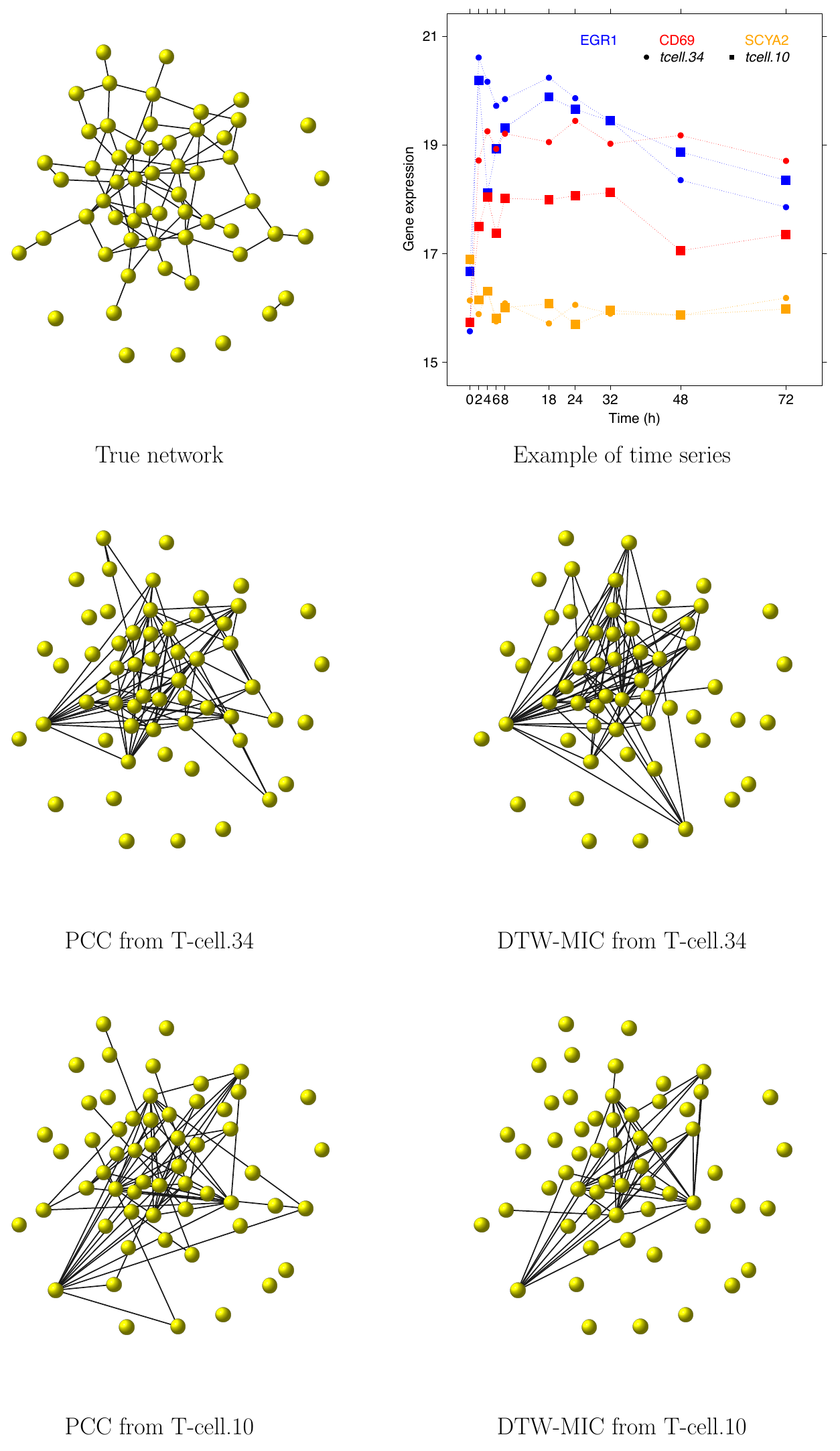}}
\end{center}
\vspace{-3cm}
\caption{\textbf{The T-cell example.} The (true) network as reconstructed by Opgen-Rhein and Strimmer (top left); the time course for three example genes EGR1 (blue), CD69 (red) and SCYA2 (orange), from replicate 1 of the \textit{tcell.34} (circles) and of the \textit{tcell.10} (squares) dataset. In the remaining panels, the networks inferred by PCC (left) and DTW-MIC from the \textit{tcell.34} (middle row) and from the \textit{tcell.10} (bottom row) dataset; in all four graphs, edges with smaller weights (threshold is 0.1 for PCC and 0.225 for DTW-MIC) are not displayed. }
\label{fig:tcell}
\end{figure}

The preprocessed array data \textit{tcell.34} and \textit{tcell.10}, log-transformed and quantile normalized, are publicly available in the R package \textit{longitudinal}.
This package was developed by Opgen-Rhein and Strimmer who inferred the corresponding network by shrinkage estimation of the (partial) dynamical correlation \cite{opgenrhein06using,opgenrhein06inferring}, which we consider here as the ground truth network, displayed in the top left panel of Fig.~\ref{fig:tcell}.
As an example of the data in the \textit{tcell.34} and \textit{tcell.10}, in the top right panel of the same Fig.~\ref{fig:tcell} we show the time course data for the three genes EGR1, CD69 and SCYA2 in the first out of 34 replicates of \textit{tcell.34} and in the first out of 10 replicates of \textit{tcell.10}.

By WGCNA, four instances of the T-cell network are inferred, by the two similarity measure PCC and DTW-MIC and starting from the two datasets \textit{tcell.34} and \textit{tcell.10}.
In both datasets, the dimension of the longitudinal data for each replicate (10 time points) cannot guarantee robustness in the inference process, since both PCC and MIC are not reliable for datasets of too small sample size \cite{gobbi13null,reshef11detecting}. 
Hence all replicates in the two datasets are consecutively joined so that time point 72h of replicate $i$ is followed by time point 0h for replicate $i+1$, thus yelding for each gene a single time course on 340 time points for \textit{tcell.34} and on 100 time points for \textit{tcell.10}.
The four inferred networks are displayed in Fig.~\ref{fig:tcell}, while in Tab.~\ref{tab:tcell} the HIM distances are reported between the true and the inferred T-cell networks.
For both datasets \textit{tcell.34} and \textit{tcell.10} the HIM distance from the true graph is smaller for the networks inferred by the DTW-MIC measure (0.16 vs. 0.21 and 0.14 vs. 024).
Note that, in all cases, the Hamming component of the distance is indeed smaller for PCC networks, while the Ipsen-Mikhailov component (and the global HIM) is larger.
Thus less links are changing between the PCC networks and the true graph, but these changing links induce a strongly different structure between the two nets, which results in larger HIM values.
Moreover, the choice of the similarity measure has a larger impact than the starting dataset, since the nets inferred using the same measure on different datasets are mutually closer than the nets inferred by different methods on the same time courses.
Finally, without the power function (with $\beta=6$ as default) applied in the WGCNA for soft thresholding the reconstructed networks are very different from the true graph, regardless of the starting dataset: the resulting HIM is about 0.47 for PCC and 0.66 for DTW-MIC, with 0.63 the average HIM value for a null model generated by computing the distance from the true graph of 1000 random network with uniform edge weight distribution in (0,1).

\section*{Conclusion}
We introduced here DTW-MIC, a novel similarity measure for inferring coexpression networks from longitudinal data as an alternative to the absolute PCC used in the WGCNA approach.
Due to the nature of its components Dynamic Time Warping and Maximal Information Coefficient, the DTW-MIC similarity can overcome the well known limitations of PCC when dealing with delayed signals and indirect interactions.
Experiments on biologically inspired synthetic data and gene expression time course data demonstrate the higher precision in the network inference achieved by DTW-MIC with respect to PCC in different conditions.

\section*{Acknowledgments}

\bibliography{riccadonna14dtw}

\begin{thebibliography}{10}
\providecommand{\url}[1]{\texttt{#1}}
\providecommand{\urlprefix}{URL }
\expandafter\ifx\csname urlstyle\endcsname\relax
  \providecommand{\doi}[1]{doi:\discretionary{}{}{}#1}\else
  \providecommand{\doi}{doi:\discretionary{}{}{}\begingroup
  \urlstyle{rm}\Url}\fi
\providecommand{\bibAnnoteFile}[1]{%
  \IfFileExists{#1}{\begin{quotation}\noindent\textsc{Key:} #1\\
  \textsc{Annotation:}\ \input{#1}\end{quotation}}{}}
\providecommand{\bibAnnote}[2]{%
  \begin{quotation}\noindent\textsc{Key:} #1\\
  \textsc{Annotation:}\ #2\end{quotation}}
\providecommand{\eprint}[2][]{\url{#2}}

\bibitem{desmet10advantages}
De~Smet R, Marchal K (2010) Advantages and limitations of current network
  inference methods.
\newblock Nature Reviews Microbiology 8: 717--729.
\bibAnnoteFile{desmet10advantages}

\bibitem{marbach10revealing}
Marbach D, Prill R, Schaffter T, Mattiussi C, Floreano D, et~al. (2010)
  Revealing strengths and weaknesses of methods for gene network inference.
\newblock PNAS 107: 6286--6291.
\bibAnnoteFile{marbach10revealing}

\bibitem{szederkenyi11inference}
Szederkenyi G, Banga J, Alonso A (2011) Inference of complex biological
  networks: distinguishability issues and optimization-based solutions.
\newblock BMC Systems Biology 5: 177.
\bibAnnoteFile{szederkenyi11inference}

\bibitem{allen12comparing}
Allen J, Xie Y, Chen M, Girard L, Xiao G (2012) {Comparing Statistical Methods
  for Constructing Large Scale Gene Networks}.
\newblock PLoS ONE 7: e29348.
\bibAnnoteFile{allen12comparing}

\bibitem{lopez13biostatistical}
L{\'o}pez-Kleine L, Leal L, L{\'o}pez C (2013) {Biostatistical approaches for
  the reconstruction of gene co-expression networks based on transcriptomic
  data}.
\newblock Briefings in Functional Genomics 12: 457--467.
\bibAnnoteFile{lopez13biostatistical}

\bibitem{rotival14leveraging}
Rotival M, Petretto E (2014) {Leveraging gene co-expression networks to
  pinpoint the regulation of complex traits and disease, with a focus on
  cardiovascular traits}.
\newblock Briefings in Functional Genomics 13: 66--78.
\bibAnnoteFile{rotival14leveraging}

\bibitem{song12comparison}
Song L, Langfelder P, Horvath S (2012) {Comparison of co-expression measures:
  mutual information, correlation, and model based indices}.
\newblock BMC Bioinformatics 13: 328.
\bibAnnoteFile{song12comparison}

\bibitem{madhamshettiwar12gene}
Madhamshettiwar P, Maetschke S, Davis M, Reverter A, Ragan M (2012) Gene
  regulatory network inference: evaluation and application to ovarian cancer
  allows the prioritization of drug targets.
\newblock Genome Medicine 4: 41.
\bibAnnoteFile{madhamshettiwar12gene}

\bibitem{baralla09inferring}
Baralla A, Mentzen W, de~la Fuente A (2009) {Inferring Gene Networks: Dream or
  Nightmare?}
\newblock Annals of the New York Academy of Science 1158: 246--256.
\bibAnnoteFile{baralla09inferring}

\bibitem{lee04coexpression}
Lee H, Hsu A, Sajdak J, Qin J, Pavlidis P (2004) {Coexpression Analysis of
  Human Genes Across Many Microarray Data Sets}.
\newblock Genome Research 14: 1085--1094.
\bibAnnoteFile{lee04coexpression}

\bibitem{lavi12network}
Lavi O, Dror G, Shamir R (2012) {Network-Induced Classification Kernels for
  Gene Expression Profile Analysis}.
\newblock Journal of Computational Biology 19: 694--709.
\bibAnnoteFile{lavi12network}

\bibitem{rapaport07classification}
Rapaport F, Zinovyev A, Dutreix M, Barillot E, Vert JP (2007) {Classification
  of microarray data using gene networks}.
\newblock BMC Bioinformatics 8: 35.
\bibAnnoteFile{rapaport07classification}

\bibitem{jansen02relating}
Jansen R, Greenbaum D, Gerstein M (2002) {Relating Whole-Genome Expression Data
  with Protein-Protein Interactions}.
\newblock Genome Research 12: 376.
\bibAnnoteFile{jansen02relating}

\bibitem{zhang05general}
Zhang B, Horvath S (2005) {A General Framework for Weighted Gene Co-Expression
  Network Analysis}.
\newblock Statistical Applications in Genetics and Molecular Biology 4: Article
  17.
\bibAnnoteFile{zhang05general}

\bibitem{langfelder08wgcna}
Langfelder P, Horvath S (2008) {WGCNA: an R package for weighted correlation
  network analysis}.
\newblock BMC Bioinformatics 9: 559.
\bibAnnoteFile{langfelder08wgcna}

\bibitem{horvath11weighted}
Horvath S (2011) {Weighted Network Analysis: Applications in Genomics and
  Systems Biology}.
\newblock Springer.
\bibAnnoteFile{horvath11weighted}

\bibitem{kumari12evaluation}
Kumari S, Nie J, Chen HS, Ma H, Stewart R, et~al. (2012) {Evaluation of Gene
  Association Methods for Coexpression Network Construction and Biological
  Knowledge Discovery}.
\newblock PLoS ONE 7: e50411.
\bibAnnoteFile{kumari12evaluation}

\bibitem{kurt14comprehensive}
Kurt Z, Aydin N, Altay G (2014) {A comprehensive comparison of association
  estimators for gene network inference algorithms}.
\newblock Bioinformatics 30: 2142--2149.
\bibAnnoteFile{kurt14comprehensive}

\bibitem{dempsey11novel}
Dempsey K, Bonasera S, Bastola D, Ali H (2011) {A Novel Correlation Networks
  Approach for the Identification of Gene Targets}.
\newblock In: Proceedings of the 44th Hawaii International Conference on System
  Sciences - HICSS 2011. IEEE, pp. 1--8.
\bibAnnoteFile{dempsey11novel}

\bibitem{reshef11detecting}
Reshef D, Reshef Y, Finucane H, Grossman S, McVean G, et~al. (2011) {Detecting
  novel associations in large datasets}.
\newblock Science 6062: 1518--1524.
\bibAnnoteFile{reshef11detecting}

\bibitem{speed11correlation}
Speed T (2011) {A Correlation for the 21st Century}.
\newblock Science 6062: 1502--1503.
\bibAnnoteFile{speed11correlation}

\bibitem{albanese12minerva}
Albanese D, Filosi M, Visintainer R, Riccadonna S, Jurman G, et~al. (2013)
  {minerva and minepy: a C engine for the MINE suite and its R, Python and
  MATLAB wrappers}.
\newblock Bioinformatics 29: 407--408.
\bibAnnoteFile{albanese12minerva}

\bibitem{nature12finding}
{Nature Biotechnology} (2012) {Finding correlations in big data}.
\newblock Nature Biotechnology 30: 334--335.
\bibAnnoteFile{nature12finding}

\bibitem{sakoe78dynamic}
Sakoe H, Chiba S (1978) {Dynamic Programming Algorithm Optimization for Spoken
  Word Recognition}.
\newblock IEEE Transactions on Acoustics, Speech, and Signal Processing 26:
  43--49.
\bibAnnoteFile{sakoe78dynamic}

\bibitem{keogh98enhanced}
Keogh E, Pazzani M (1998) {An enhanced representation of time series which
  allows fast and accurate classification, clustering and relevance feedback}.
\newblock In: Press A, editor, Proc. KDD '98. pp. 239--241.
\bibAnnoteFile{keogh98enhanced}

\bibitem{keogh00scaling}
Keogh E, Pazzani M (2000) {Scaling up dynamic time warping for datamining
  applications}.
\newblock In: Press A, editor, Proc. KDD '00. pp. 285--289.
\bibAnnoteFile{keogh00scaling}

\bibitem{aach01aligning}
Aach J, Church G (2001) {Aligning gene expression time series with time warping
  algorithms}.
\newblock Bioinformatics 17: 495--508.
\bibAnnoteFile{aach01aligning}

\bibitem{furlanello06combining}
Furlanello C, Merler S, Jurman G (2006) {Combining feature selection and DTW
  for time-varying functional genomics}.
\newblock IEEE Transactions on Signal Processing 54: 2436--2443.
\bibAnnoteFile{furlanello06combining}

\bibitem{keogh01derivative}
Keogh E, Pazzani M (2001) {Derivative Dynamic Time Warping}.
\newblock In: Kumar V, Grossman R, editors, Proceedings of the 2001 SIAM
  International Conference on Data Mining. pp. 1--11.
\bibAnnoteFile{keogh01derivative}

\bibitem{rakthanmanon12searching}
Rakthanmanon T, Campana B, Mueen A, Westover G, Zhu Q, et~al. (2012) {Searching
  and mining trillions of time series subsequences under Dynamic Time Warping}.
\newblock In: Q Y, Agarwal D, Pei J, editors, Proceedings of the 18th ACM
  SIGKDD International Conference on Knowledge Discovery and Data Mining
  KDD'12. pp. 262--270.
\bibAnnoteFile{rakthanmanon12searching}

\bibitem{batista14cid}
Batista G, Keogh E, Tatawi O, de~Souza V (2014) {CID: an efficient
  complexity-invariant distance for time series}.
\newblock Data Mining and Knowledge Discovery 28: 634--669.
\bibAnnoteFile{batista14cid}

\bibitem{li14online}
Li H (2014) {On-line and dynamic time warping for time series data mining}.
\newblock International Journal of Machine Learning and Cybernetics April: 1-9.
\bibAnnoteFile{li14online}

\bibitem{elbakry10inference}
ElBakry O, Ahmad M, Swamy M (2010) Inference of gene regulatory networks from
  time-series microarray data.
\newblock In: Proc. 8th IEEE NEWCAS Conference. IEEE, pp. 141--144.
\bibAnnoteFile{elbakry10inference}

\bibitem{filosi14stability}
Filosi M, Visintainer R, Riccadonna S, Jurman G, Furlanello C (2014) {Stability
  Indicators in Network Reconstruction}.
\newblock PLoS ONE 9: e89815.
\bibAnnoteFile{filosi14stability}

\bibitem{jurman14him}
Jurman G, Visintainer R, Riccadonna S, Filosi M, Furlanello C (2014) {The HIM
  glocal metric and kernel for network comparison and classification}.
\newblock ArXiv:1201.2931v3 [math.CO].
\bibAnnoteFile{jurman14him}

\bibitem{shaffter11genenetweaver}
Schaffter T, Marbach D, Floreano D (2011) {Gene{N}et{W}eaver: {I}n silico
  benchmark generation and performance profiling of network inference methods}.
\newblock Bioinformatics 27: 2263--2270.
\bibAnnoteFile{shaffter11genenetweaver}

\bibitem{prill10towards}
Prill R, Marbach D, Saez-Rodriguez J, Sorger P, Alexopoulos L, et~al. (2010)
  {Towards a Rigorous Assessment of Systems Biology Models: The DREAM3
  Challenges}.
\newblock PLoS ONE 5: e9202.
\bibAnnoteFile{prill10towards}

\bibitem{rangel04modeling}
Rangel C, Angus J, Ghahramani Z, Lioumi M, Sotheran E, et~al. (2004) Modeling
  t-cell activation using gene expression profiling and state-space models.
\newblock Bioinformatics 20: 1361--1372.
\bibAnnoteFile{rangel04modeling}

\bibitem{filosi14renette}
Filosi M, Droghetti S, Arbitrio E, Visintainer R, Riccadonna S, et~al. (2014)
  {ReNette: a web-service for network reproducibility analysis}.
\newblock {bioRxiv}-doi:10.1101/008433, submitted.
\bibAnnoteFile{filosi14renette}

\bibitem{r14manual}
{R Core Team} (2014) R: A Language and Environment for Statistical Computing.
\newblock R Foundation for Statistical Computing, Vienna, Austria.
\newblock \urlprefix\url{http://www.R-project.org/}.
\bibAnnoteFile{r14manual}

\bibitem{gusfield97algorithms}
Gusfield D (1997) Algorithms on strings, trees and sequences.
\newblock Cambridge University Press.
\bibAnnoteFile{gusfield97algorithms}

\bibitem{giorgino09computing}
Giorgino T (2009) {Computing and Visualizing Dynamic Time Warping Alignments in
  R: The dtw Package}.
\newblock Journal of Statistical Software 31: 1--24.
\bibAnnoteFile{giorgino09computing}

\bibitem{desollaprice65networks}
de~Solla~Price D (1965) {Networks of Scientific Papers}.
\newblock Science 149: 510--515.
\bibAnnoteFile{desollaprice65networks}

\bibitem{barabasi99emergence}
Barabasi AL, Albert R (1999) {Emergence of scaling in random networks}.
\newblock Science 286: 509-–512.
\bibAnnoteFile{barabasi99emergence}

\bibitem{jurman10introduction}
Jurman G, Visintainer R, Furlanello C (2011) {An introduction to spectral
  distances in networks}.
\newblock Frontiers in Artificial Intelligence and Applications 226: 227--234.
\bibAnnoteFile{jurman10introduction}

\bibitem{tun06metabolic}
Tun K, Dhar P, Palumbo M, Giuliani A (2006) Metabolic pathways variability and
  sequence/networks comparisons.
\newblock BMC Bioinformatics 7: 24.
\bibAnnoteFile{tun06metabolic}

\bibitem{dougherty10validation}
Dougherty E (2010) Validation of gene regulatory networks: scientific and
  inferential.
\newblock Briefings in Bioinformatics 12: 245--252.
\bibAnnoteFile{dougherty10validation}

\bibitem{ipsen02evolutionary}
Ipsen M, Mikhailov A (2002) Evolutionary reconstruction of networks.
\newblock Phys Rev E 66: 046109.
\bibAnnoteFile{ipsen02evolutionary}

\bibitem{marbach09generating}
Marbach D, Schaffter T, Mattiussi C, Floreano D (2009) {Generating {R}ealistic
  {I}n {S}ilico {G}ene {N}etworks for {P}erformance {A}ssessment of {R}everse
  {E}ngineering {M}ethods}.
\newblock Journal of {C}omputational {B}iology 16: 229--239.
\bibAnnoteFile{marbach09generating}

\bibitem{shaffter12gnw}
Shaffter T, Marbach D, Roulet G (2012) {GeneNetWeaver User Manual, version
  3.1}.
\newblock
  \url{http://tschaffter.ch/projects/gnw/downloads/3.1b/gnw-3.1b-user-manual.pdf}.
\bibAnnoteFile{shaffter12gnw}

\bibitem{stolovitzky07dialogue}
Stolovitzky G, Monroe D, Califano A (2007) {Dialogue on Reverse-Engineering
  Assessment and Methods}.
\newblock Annals of the New York Academy of Sciences 1115: 1--22.
\bibAnnoteFile{stolovitzky07dialogue}

\bibitem{stolovitzky09lessons}
Stolovitzky G, Prill R, Califano A (2009) {Lessons from the DREAM2 Challenges}.
\newblock Annals of the New York Academy of Sciences 1158: 159--195.
\bibAnnoteFile{stolovitzky09lessons}

\bibitem{prill11crowdsourcing}
Prill R, Saez-Rodriguez J, Alexopoulos L, Sorger P, Stolovitzky G (2011)
  {Crowdsourcing Network Inference: The DREAM Predictive Signaling Network
  Challenge}.
\newblock Science Signaling 4: mr7.
\bibAnnoteFile{prill11crowdsourcing}

\bibitem{marbach12wisdom}
Marbach D, Costello J, Kuffner R, Vega N, Prill R, et~al. (2012) {Wisdom of
  crowds for robust gene network inference}.
\newblock Nature Methods 9: 796--804.
\bibAnnoteFile{marbach12wisdom}

\bibitem{balaji06comprehensive}
Balaji S, Babu M, Iyer L, Luscombe N, Aravind L (2006) {Comprehensive Analysis
  of Combinatorial Regulation using the Transcriptional Regulatory Network of
  Yeast}.
\newblock Journal of Molecular Biology 360: 213--227.
\bibAnnoteFile{balaji06comprehensive}

\bibitem{gamacastro08regulondb}
Gama-Castro S, Jiménez-Jacinto V, Peralta-Gil M, Santos-Zavaleta A,
  Peñaloza-Spinola M, et~al. (2008) {RegulonDB (version 6.0): gene regulation
  model of Escherichia coli K-12 beyond transcription, active (experimental)
  annotated promoters and Textpresso navigation}.
\newblock Nucleic Acids Research 36: D120--D124.
\bibAnnoteFile{gamacastro08regulondb}

\bibitem{tu02quantitative}
Tu Y, Stolovitzky G, Klein U (2002) {Quantitative noise analysis for gene
  expression microarray experiments}.
\newblock Proceedings of the National Academy of Sciences 99: 14031--14036.
\bibAnnoteFile{tu02quantitative}

\bibitem{opgenrhein06using}
Opgen-Rhein R, Strimmer K (2006) {Using regularized dynamic correlation to
  infer gene dependency networks from time-series microarray data}.
\newblock In: Ruusuvuori P, Manninen T, Huttunen H, Linne ML, Yli-Harja O,
  editors, Proc. 4th International WCSB 2006. pp. 73--76.
\bibAnnoteFile{opgenrhein06using}

\bibitem{opgenrhein06inferring}
Opgen-Rhein R, Strimmer K (2006) {Inferring gene dependency networks from
  genomic longitudinal data: a functional data approach}.
\newblock REVSTAT 4: 53--65.
\bibAnnoteFile{opgenrhein06inferring}

\bibitem{gobbi13null}
Gobbi A, Jurman G (2013) {A null model for Pearson correlation networks}.
\newblock {bioRxiv}-doi:10.1101/001065, submitted.
\bibAnnoteFile{gobbi13null}

\end{thebibliography}

\cleardoublepage

\section*{Figure Legends}

\cleardoublepage

\cleardoublepage

\cleardoublepage

\cleardoublepage

\cleardoublepage

\cleardoublepage

\cleardoublepage

\cleardoublepage

\cleardoublepage

\section*{Tables}

\begin{table}[!ht]
\caption{HIM distances with basic statistics of the DTW-MIC (D) and the PCC (P) inferred networks for all experiments on the GNW datasets $\textrm{Yeast}_{20}, \textrm{Ecoli}_{20}, \textrm{Ecoli}_{50}$}
\label{tab:gnw}
\begin{center}
\begin{tabular}{r|rr|rr|rr}
\# Dataset &\multicolumn{2}{r}{$\textrm{Yeast}_{20}$} & \multicolumn{2}{c}{$\textrm{Ecoli}_{20}$} & \multicolumn{2}{c}{$\textrm{Ecoli}_{50}$}\\
& P & D & P & D & P & D \\ 
\hline                               
$d_1$ & 0.57 & 0.23 & 0.37 & 0.20&  0.22 & 0.22 \\ 
$d_2$ & 0.41 & 0.21 & 0.41 & 0.19&  0.31 & 0.20 \\ 
$d_3$ & 0.39 & 0.20 & 0.37 & 0.19&  0.23 & 0.23 \\ 
$d_4$ & 0.25 & 0.25 & 0.23 & 0.36&  0.27 & 0.21 \\ 
$d_5$ & 0.56 & 0.24 & 0.41 & 0.19&  0.35 & 0.19 \\ 
$d_6$ & 0.35 & 0.19 & 0.53 & 0.20&  0.40 & 0.20 \\ 
$d_7$ & 0.56 & 0.23 & 0.40 & 0.19&  0.26 & 0.21 \\ 
$d_8$ & 0.42 & 0.22 & 0.52 & 0.20&  0.29 & 0.21 \\ 
$d_9$ & 0.49 & 0.22 & 0.42 & 0.20&  0.25 & 0.21 \\
$d_{10}$ &0.53  & 0.22 & 0.22 & 0.28&  0.35 & 0.20 \\
\hline
Mean    & 0.45 & 0.22 & 0.39 & 0.22 & 0.29 & 0.21 \\ 
Median  & 0.45 & 0.22 & 0.40 & 0.20 &  0.28 & 0.21\\ 
Std. Dev. & 0.10 & 0.02 & 0.10 & 0.06 & 0.06 & 0.01\\
\end{tabular} 
\begin{flushleft}
\end{flushleft}
\end{center}
\end{table}

\cleardoublepage

\begin{table}[!ht]
\caption{Hamming (H), Ipsen-Mikhailov (IM) and HIM distances among the true (TN) and the T-cell WGCNA inferred networks, by PCC (P) and DTW-MIC (D) similarity measure, from the \textit{tcell.34} (34) and the \textit{tcell.10} (10) time course datasets.}
\label{tab:tcell}
\begin{center}
\begin{tabular}{ll|rrr}
Net1 & Net2 & H & IM & HIM\\
\hline
TN  &P\_34   & 0.060 & 0.296 & 0.214 \\
TN  &D\_34   & 0.112 & 0.203 & 0.164 \\
TN  &P\_10   & 0.059 & 0.347 & 0.249 \\
TN  &D\_10   & 0.094 & 0.176 & 0.141 \\
P\_10&P\_34   & 0.022 & 0.058 & 0.044 \\
D\_10&D\_34   & 0.055 & 0.171 & 0.127 \\
P\_34&D\_34   & 0.064 & 0.493 & 0.351 \\
P\_10&D\_10   & 0.043 & 0.474 & 0.336 \\
\end{tabular} 
\begin{flushleft}
\end{flushleft}
\end{center}
\end{table}

\end{document}